\documentclass[%
superscriptaddress,
 amsmath,amssymb,
 aps,
 prd,
floatfix,nofootinbib
]{revtex4-1}

\usepackage{graphicx}
\usepackage{dcolumn}
\usepackage{bm}
\usepackage[normalem]{ulem}
\usepackage{tikz}
\usetikzlibrary{positioning}

\usepackage{braket}
\usepackage{multirow}
\usepackage{dsfont}
\usepackage{mathtools}
\usepackage{xcolor}
\usepackage[normalem]{ulem}

\usepackage{cancel}
\usepackage{epsfig,dsfont}
\usepackage{subfigure}
\usepackage{bm}
\usepackage{amssymb}
\usepackage{mathrsfs}
\usepackage{amsmath}
\usepackage{enumitem}
\usepackage{dcolumn}

\begin{document}

\title{Fluctuations of conserved charges, chiral spin symmetry and
deconfinement in an $SU(2)_{color}$ subgroup of $SU(3)_{color}$ above $T_c$.}

\author{L.Ya.~Glozman}
\affiliation{Institute of Physics, University of Graz, 8010 Graz, Austria}

\date{\today}

\begin{abstract}
\vspace{4mm}
Above a pseudocritical temperature of chiral symmetry
restoration $T_c$ the energy and the pressure are very far from the
quark-gluon-plasma limit (i.e. ideal gas of free quarks and gluons). At the same time very soon above $T_c$ fluctuations of
conserved charges behave as if quarks were free particles. Within
the $T_c - 3T_c$ interval a chiral spin symmetry emerges in QCD which
is not consistent with free quarks and suggests that degrees
of freedom are chirally symmetric quarks bound into the color-singlet
objects by the chromoelectric field. Here we analyse temporal
and spatial correlators in this interval and demonstrate that they
indicate simultaneously  the chiral spin symmetry as well as
absence of the interquark interactions in  channels constrained
by a current conservation. The latter channels are responsible
for both fluctuations of conserved charges and for dileptons.
Assuming that a $SU(2)_{color}$ subgroup of $SU(3)_{color}$ is deconfined
soon above $T_c$ but confinement  persits in $SU(3)_{color}/SU(2)_{color}$  in the
interval $T_c - 3T_c$  we are able
to reconcile all empirical facts listed above.
\end{abstract}
\maketitle


\section{\label{sec:intro}Introduction}

It is understood that QCD at low temperatures and vanishing baryon chemical potential is a hadron (meson) gas. At temperatures between 100 MeV
and 200 MeV a very smooth chiral symmetry restoration crossover is observed
and a temerature $T_c \sim 155$ MeV could be approximately considered
as a pseudocritical temperature of chiral symmetry restoration \cite{F1,K}. 
There is no definition and order parameter
for deconfinement in QCD with light quarks except that deconfinement
should be accompanied by a free motion of colored quarks and gluons
within a macroscopical piece of matter. Such a situation should take place
at very high temperatures where due to asymptotic freedom the strong
interaction coupling vanishes.
What the strongly interacting matter is 
above the chiral symmetry restoration, is a big puzzle. It cannot be
a quark-gluon-plasma (QGP) that is defined as a gas of free (i.e. deconfined)
quarks and gluons.  For example, a perfect fluidity of the QCD matter above $T_c$  is
not consistent with free  quarks and gluons as degrees of freedom.
Another evidence that QCD is a strongly interacting matter above $T_c$
is its pressure and energy density  that are
very far from the Stefan-Boltzmann limit, i.e. the limit of free noninteracting
quarks and gluons, above $T_c$ but below $3T_c$ \cite{eq1,eq2}. 

\begin{figure}
  \centering
  \includegraphics[scale=0.45]{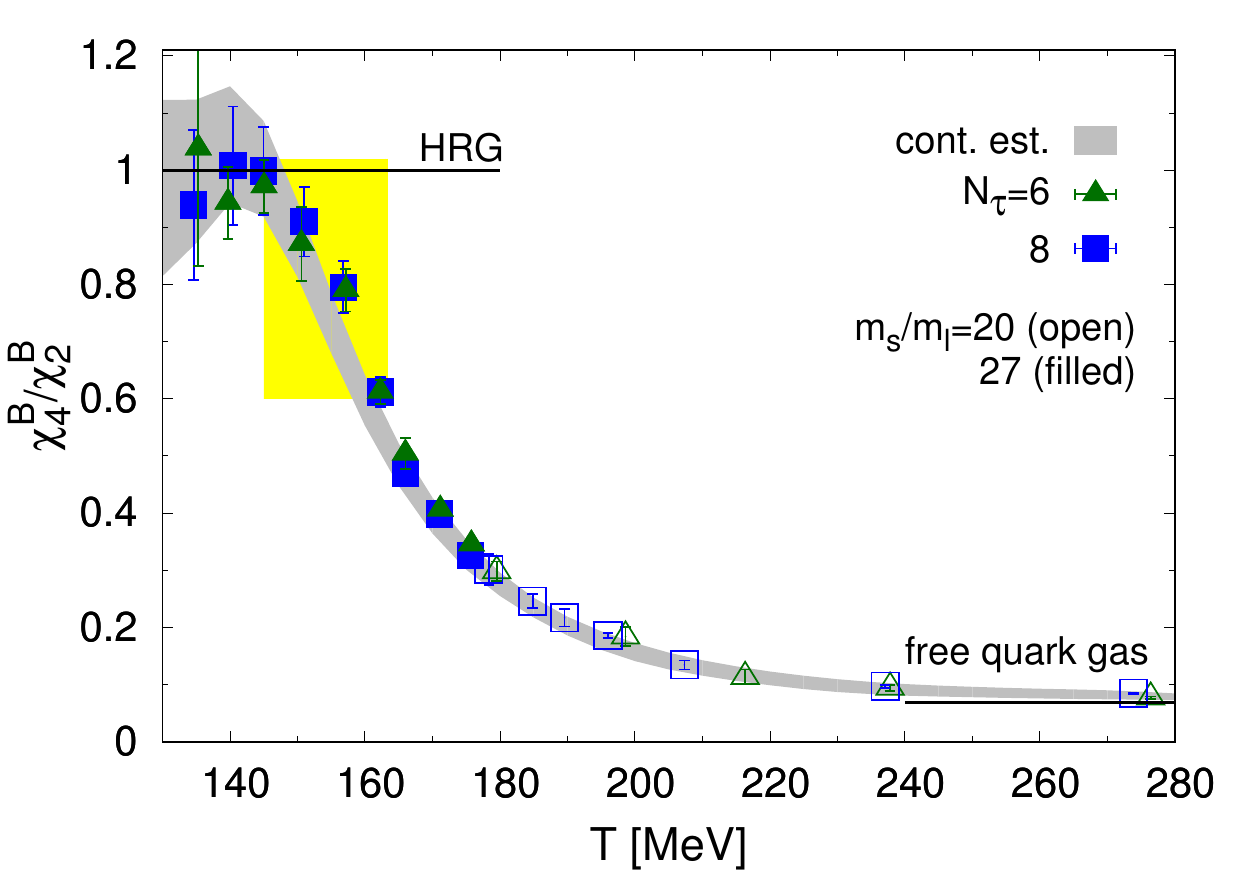} 
  \includegraphics[scale=0.45]{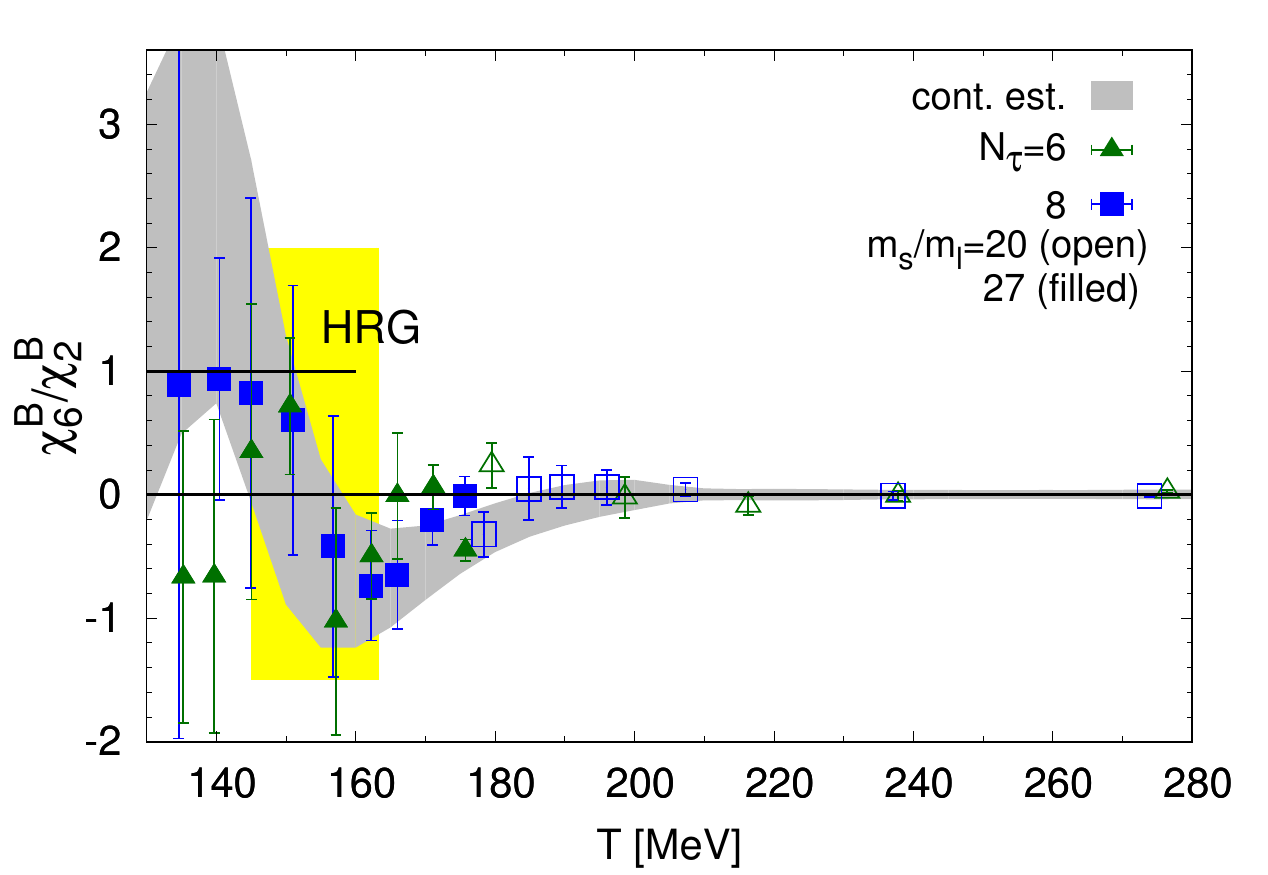} 
  \caption{ Left: The ratio of fourth and second order cumulants of net-baryon number fluctuations  versus temperature.
Right: same as the left hand side, but for the ratio of sixth and second order cumulants of net-baryon number fluctuations
B. The boxes indicate the transition region, $T_c = (154 \pm 9)$ MeV. Grey bands show continuum estimate. From ref. \cite{Bazavov}.
}
\label{fl}
\end{figure}

At the same time fluctuations of conserved charges \cite{Asakawa,Koch,Ejiri,Bazavov}
approach the limit of noninteracting quarks
very soon after $T_c$, see Fig. \ref{fl}.
This behavior of fluctuations is considered sometimes as a signal of deconfinement. But how  to reconcile then this with pressure and energy
density as well as with perfect fluidity?

It was recently found on the lattice \cite{R1,R2,R3}  that temporal and spatial
correlators of $N_F=2$ QCD reveal above $T_c$ but below $3 T_c$ multiplet patterns of  chiral spin  $SU(2)_{CS}$ and $SU(2N_F)$ groups \cite{G1,G2}.
These groups are not symmetries of the Dirac action and hence inconsistent
with free (deconfined) quarks. At the same time they are symmetries
of the Lorentz-invariant color charge and of the chromoelectric interaction
in a given reference frame. Observation of these symmetries suggests that
in the $T_c - 3 T_c$ interval degrees of freedom are chirally symmetric
quarks bound into color-singlet objects by the chromoelectric field and
effects of the chromomagnetic field are at least strongly suppressed.
The chemical potential term in the QCD action is manifestly chiral spin
symmetric which means that this symmetry should persist at a finite chemical potential either \cite{G3}.  These symmetries prohibit a finite axial chemical potential
and consequently an electric current in QCD matter induced by an
external magnetic field should be either absent or very small \cite{G4}.
The experimental search of the chiral magnetic effect \cite{CME1,CME2} 
at RHIC and LHC hints that indeed the magnetically induced
electric current either vanishes or very
small \cite{CME}. If confirmed, it would be an experimental verification of
emerged chiral spin symmetry above $T_c$.

A very strange and self-contradictory picture emerges. On the one hand
there are signals that quarks are deconfined soon above $T_c$, as
it follows from fluctuations of conserved charges and from absence of
$\rho$-like bound states above $T_c$ seen via dilepton production. At the same time
quarks cannot be deconfined because of very clear chiral spin symmetry
in QCD above $T_c$ as well as because of the energy density and pressure
and perfect fluidity properties. In this paper we reanalyse existing correlators
\cite{R1,R2,R3} and demonstrate that while correlators show clear
patterns of the chiral spin symmetry, they are simultaneously consistent with
 absence of the interquark
interactions in channels constrained by a conserved  current.
Hence these correlators are compatible with known results
about fluctuations of conserved charges and absence of bound states as
seen by dilepton production. This suggests that we
have an evidence of confinement of quarks and at the same time
their deconfined-like behavior in channels controlled by a conserved  
current.
This paradox  could be explained by the assumption that
while deconfinement, i.e. a screening of the color charge happens
above $T_c$ in the  $SU(2)_{color}$ subgroup of $SU(3)_{color}$, it does
not happen in the $SU(3)_{color}/SU(2)_{color}$. This way we automatically
obtain the chiral spin symmetry of correlators and at same time
absence of interactions in channels constrained by a conserved current.
This assumption allows to reconcile all empirical facts listed above.

\section{Chiral spin symmetry as a symmetry of the color charge}

The $SU(2)_{CS}$ chiral spin (CS) transformations \cite{G1,G2} are defined by
\begin{equation}
\psi(x) \; \rightarrow \; \exp\left(\frac{i}{2}\vec\Sigma \, \vec\epsilon\right)\psi(x) \; , \quad \;
\bar{\psi}(x) \; \rightarrow \; \bar{\psi}(x) \gamma_4 \exp\left(-\frac{i}{2}\vec\Sigma \, \vec\epsilon\right) \gamma_4 \; ,
\label{equ:su2cstrafos}
\end{equation}
where $\vec\epsilon$ are the rotation parameters\footnote{This symmetry
was reconstructed from a hadron degeneracy observed on the lattice
upon artificial truncation of the near-zero modes of the Dirac operator \cite{D1,D2}.}. For the generators $\vec\Sigma$ there are 
four different choices $\vec\Sigma = \vec\Sigma_k$ with $k = 1,2,3,4$.
The choice of $k$ for a given observable is constrained by the
rotational $O(3)$ symmetry. Namely, the CS transformations should
not mix operators with different angular momentum.
 The generators are given by
\begin{align}
\vec\Sigma_k \; = \; \{\gamma_k,-i\gamma_5\gamma_k,\gamma_5\} \; ,
\end{align}
and the $su(2)$ algebra is  satisfied for any choice $k=1,2,3,4$.
Here  we use the set $\gamma_\mu, \mu = 1,2,3,4$
of hermitian Euclidean $\gamma$-matrices that fulfill the anti-commutation relations
\begin{equation}
\gamma_\mu \gamma_\nu + \gamma_\nu \gamma_\mu \; = \; 
2\delta_{\mu \nu}\; , \; \qquad \gamma_5 \; \equiv \; \gamma_1\gamma_2\gamma_3\gamma_4 \; .
\label{eq:diracalgebra}
\end{equation}

The CS symmetry is not a symmetry of the Dirac Lagrangian,
hence it cannot exist for free deconfined quarks.
At the same time it is a symmetry of the Lorentz-invariant 
color charge in QCD

\begin{equation}
Q_c^a \;  = \; \int \!\! d^3x  \; 
\psi^\dagger(x) \frac{t^a}{2} \psi(x), ~~~ a = 1,2,...,N^2_c -1,
\label{Qc}
\end{equation}
where $t^a$ are the color $SU(N_c)$ generators. The color charge
is invariant under a unitary transformation that acts only in the
Dirac space. The gluon part of the color charge is automatically
CS-invariant.

This symmetry of the color charge has  important implications.
Namely, the chromoelectric field in a given reference frame
is defined via its action on the probe color charge. Since the
color charge is invariant under the CS transformations, the
chromoelectric interaction in a given reference frame is also
CS - invariant. The chromomagnetic field is defined via its
action on the color spatial current 
$ \bar{\psi} \gamma^i \frac{t^a}{2} \psi$. The latter current is not CS-invariant.
Hence the chromomagnetic and chromoelectric interactions
in a given reference frame can be distinguished by the CS symmetry.
Obviously, to discuss the chromoelectric and chromomagnetic
parts of gluonic field one needs to fix a reference frame since
they transform through each other upon a Lorentz transformation.

This can be made more explicit.
In Minkowski space in a given reference frame the
chromoelectric and chromomagnetic fields are different fields.
The quark-gluon interaction Lagrangian can be split into temporal and spatial parts:
\begin{equation}
\overline{\psi} \,  \gamma^{\mu} D_{\mu} \, \psi  \; = \; \overline{\psi}  \, \gamma^0 D_0  \, \psi 
\; + \; \overline{\psi} \,  \gamma^i D_i \, \psi \; ,
\label{cl}
\end{equation}
where 
\begin{equation}
D_{\mu}\psi \; = \; \left( \partial_\mu - ig \frac{{ t} \cdot {\bf A}_\mu}{2}\right )\psi \; .
\end{equation}
The temporal term includes the interaction of the color-octet charge density 
\begin{equation}
\bar \psi (x) \,  \gamma^0  \frac{{ t^a}}{2} \, \psi(x) \; = \; \psi (x)^\dagger \, \frac{ t^a }{2} \, \psi(x)
\label{den}
\end{equation}
with the chromo-electric component  $A^a_0$ of the gluonic field. It is invariant  
under $SU(2)_{CS}$. We stress that the $SU(2)_{CS}$ transformations
defined in Eq.~(\ref{equ:su2cstrafos}) via the Euclidean
Dirac matrices can be identically applied to Minkowski Dirac spinors without
any modification of the generators.
The spatial part contains the quark kinetic term and the interaction with the chromomagnetic field. This term 
breaks $SU(2)_{CS}$. In other words: the $SU(2)_{CS}$ symmetry distinguishes between quarks interacting 
with the chromoelectric and chromomagnetic components of the gauge field. We note that  
discussing ``electric'' and ``magnetic'' components can be done
only in Minkowski space and in addition one needs to fix a reference frame.  At high 
temperatures Lorentz invariance is broken and a natural frame to discuss physics is the rest frame of the medium. 

The group $SU(2)_{CS} \otimes SU(N)_F$ can be extended to $SU(2N_F)$ with  generators:
\begin{align}
\{ (\vec{\tau} \otimes \mathds{1}_D) ,
(\mathds{1}_F \otimes \vec \Sigma_k) ,
(\vec{\tau} \otimes \vec \Sigma_k) \} \; .
\label{gensSU4}
\end{align} 
The corresponding transformations are a straightforward generalization of Eq.~(\ref{equ:su2cstrafos}) 
obtained by replacing the generators 
$\vec{\Sigma}$ by those listed in (\ref{gensSU4}).  Also the  group $SU(2N_F)$ is a symmetry of  the
color charge (\ref{Qc})
as well as of the quark - chromoelectric
interaction terms of the QCD Lagrangian.

\section{Review of the temporal correlators above $T_c$}

\begin{figure}
\centering
\includegraphics[angle=0,width=0.45\linewidth]{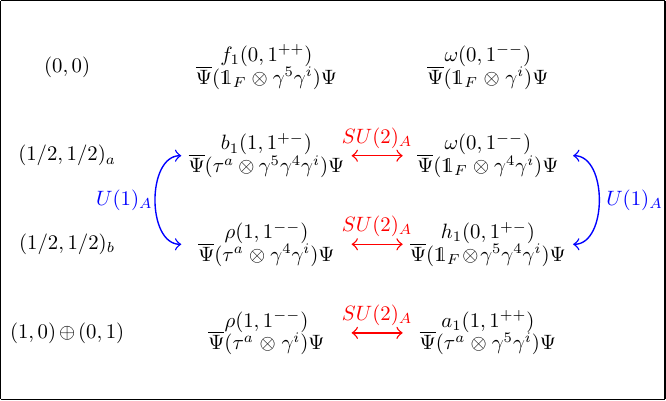}
\includegraphics[angle=0,width=0.45\linewidth]{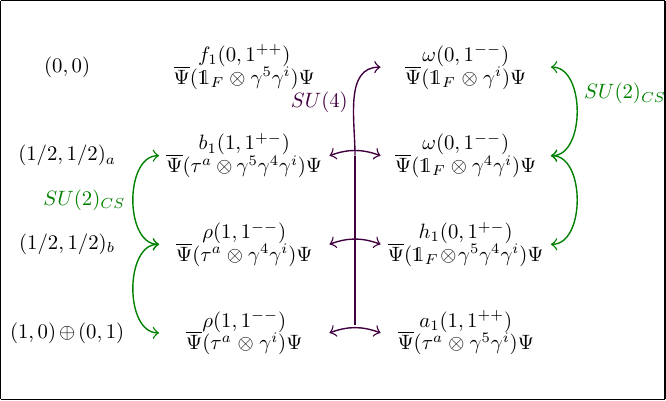}
\caption{Transformations between $J=1$ operators, $i=1,2,3$.
The left columns indicate the $SU(2)_L \times SU(2)_R$ 
representation for every
operator. Red and blue arrows connect operators which transform into 
each other under $SU(2)_L \times SU(2)_R$ and $U(1)_A$, respectively.
Green arrows connect operators that belong to
$SU(2)_{CS}$, $k=4$  triplets. Purple arrow shows the $SU(4)$
15-plet. The $f_1$ operator is is a singlet of $SU(4)$.
From Ref. \cite{G2}.}
\label{F1}
\end{figure}

The $SU(2)_L \times SU(2)_R$ and $U(1)_A$ transformation
properties of the $J=1$ operators, relevant for temporal correlators, are given in the left panel of Fig. \ref{F1}
while their $SU(2)_{CS}$ and $SU(4)$ multiplets are presented in the
right panel of this figure \cite{G2}.

\begin{figure}
  \centering
  \includegraphics[scale=0.6]{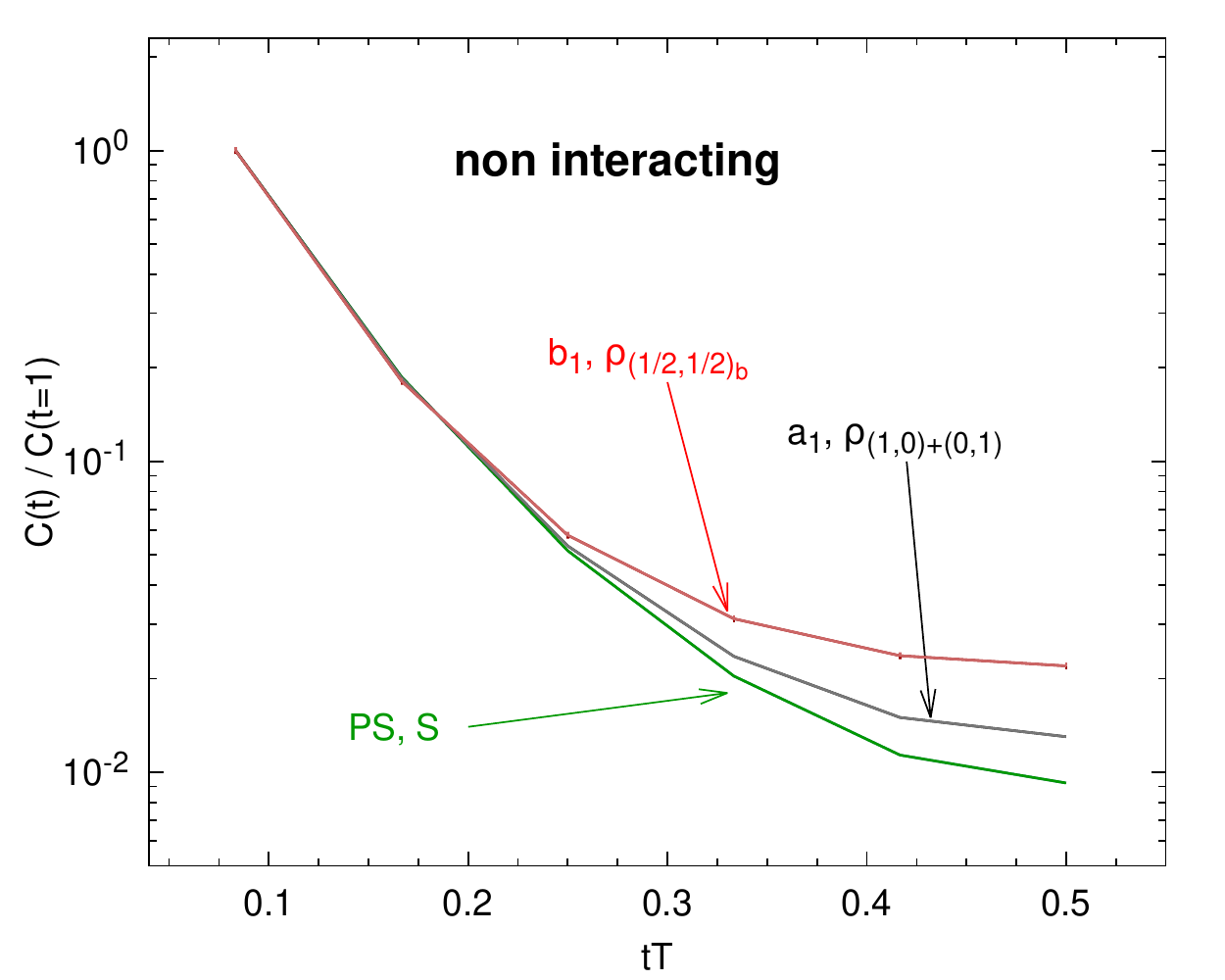} 
  \includegraphics[scale=0.6]{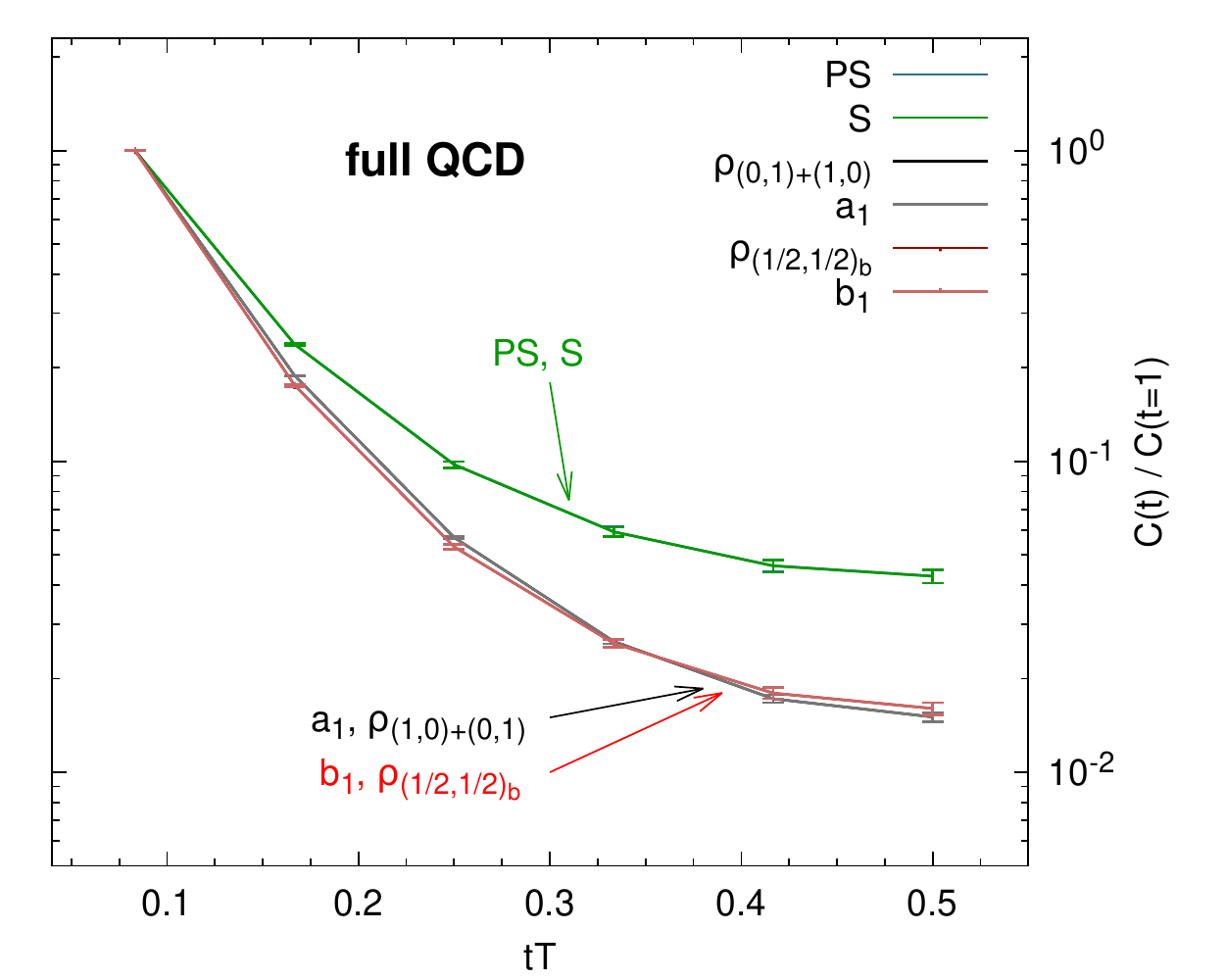} 
\caption{ Temporal correlation functions for $12 \times 48^3$
lattices. The l.h.s. shows correlators calculated with free
noninteracting quarks with manifest $U(1)_A$  and $SU(2)_L \times SU(2)_R$
symmetries. The r.h.s. presents full QCD results at a temperature $ T=220$ MeV ($1.2 T_c$),
which shows multiplets of all  $U(1)_A$, $SU(2)_L \times SU(2)_R$, $SU(2)_{CS}$  and $SU(4)$ groups. From Ref. \cite{R3}.
}
\label{tcorr}
\end{figure}

On the r.h.s.  of Fig.~\ref{tcorr} we show temporal correlators

\begin{equation}
C_\Gamma(t) = \sum\limits_{x, y, z}
\braket{\mathcal{O}_\Gamma(x,y,z,t)
\mathcal{O}_\Gamma(\mathbf{0},0)^\dagger},
\label{eq:momentumprojection}
\end{equation}
at a temperature $T = 220$ MeV ($1.2 T_c$)  calculated in $N_F=2$
QCD with a chirally symmetric Domain Wall Dirac operator 
with physical quark masses  \cite{R3}.
Here $\mathcal{O}_\Gamma(x,y,z,t)$ is an operator that creates a
quark-antiquark pair  with fixed quantum numbers. Summation over
$x,y,z$ projects out the  rest frame. 

Above the chiral restoration crossover
we apriori expect  in observables the chiral 
$SU(2)_L \times SU(2)_R$ symmetry.   This symmetry 
is evidenced by degeneracy of correlators of the
operators $\rho_{(0,1)+(1.0)}$ and $a_1$ connected by the
$SU(2)_L \times SU(2)_R$ transformation.
While the axial anomaly is a pertinent
property of QCD its effect is determined by the topological charge
density. There are strong evidences from the lattice that the
$U(1)_A$ symmetry is also effectively restored above $T_c$ \cite{Tomiya:2016jwr}.
 The
$U(1)_A$ restoration is signalled by  degeneracy of correlators 
of the isovector scalar (S) and isovector pseudoscalar (PS)
operators connected by the
$U(1)_A$ transformation. It is also evidenced by degeneracy
of the correlators with $b_1$ and $\rho_{(1/2,1/2)_b}$ operators.

 An approximate degeneracy of the  $b_1$, $\rho_{(1,0)+(0,1)}$
and  $\rho_{(1/2,1/2)_b}$ correlators indicates emergent $SU(2)_{CS}$  
symmetry since these three operators form a CS triplet. Their breaking is estimated at the level of less than $ 5\%$. A degeneracy of all four
$J=1$ correlators $b_1$, $\rho_{(1,0)+(0,1)}$,
$\rho_{(1/2,1/2)_b}$ and $a_1$
on the r.h.s.  of Fig.~\ref{tcorr} evidences emergence
of the $SU(4)$ symmetry.

On the l.h.s of Fig.~\ref{tcorr} we present correlators
calculated with noninteracting quarks on the same lattice.
They represent a QGP at a very high temperature. 
Free quarks are governed by the Dirac
equation and only $U(1)_A$ and $SU(2)_L \times SU(2)_R$ chiral symmetries
exist. Indeed, here we observe exact degeneracy of all correlators
connected by $U(1)_A$ and $SU(2)_L \times SU(2)_R$ transformations.

A qualitative difference between the pattern on the l.h.s.   and the pattern on the r.h.s of Fig.~\ref{tcorr}
is obvious.  While for free (i.e., deconfined quarks) on the 
l.h.s of Fig.~\ref{tcorr} we observe multiplets of $U(1)_A$ and $SU(2)_L \times SU(2)_R$ groups, on the r.h.s. of the same figure we clearly see
multiplets of all  $U(1)_A$, $SU(2)_L \times SU(2)_R$, $SU(2)_{CS}$  and $SU(4)$ groups.

The $SU(2)_{CS}$ and $SU(4)$ groups are not symmetries of the Dirac
action and hence they are incompatible with free deconfined quarks.
Since these are symmetries of the color charge in QCD and of the chromoelectric
interaction, while the chromomagnetic interaction breaks them, we conclude
that elementary objects are the chirally symmetric quarks bound
by the chromoelectric field into the color-singlet objects and a
contribution of the chromomagnetic field is at least very
strongly suppressed.

All these properties have been discussed  in Ref. \cite{R3}.
Now we add a very important observation. While
correlators of most operators  are very different as compared
to the noninteracting quark system, this is not true for the correlators
of the $\rho_{(1,0)+(0,1)}$ operator. The correlators of the latter
operator on the r.h.s and on the l.h.s. are very close.
The vector $\rho_{(1,0)+(0,1)}$ operator is constrained by the
 current conservation:
 
\begin{equation}
\partial_\mu (\bar \psi \frac{\tau^a}{2} \gamma^\mu \psi) = 0. 
\label{current} 
\end{equation} 
  The current 
conservation is a consequence of a global $SU(2)_V$ isospin symmetry.
While in general
the axial vector current $a_1$ is not constrained by the current conservation,
it is in the chirally symmetric regime and correlators of the vector
and axial vector operators are identical.

These results suggest, that while the interquark interaction is strong
and the system in QCD above $T_c$ is not a system of free quarks
(the latter would require that  correlators in QCD and for free
quarks  were identical for all possible quantum numbers)
the interquark interaction in channels that are constrained by the
current conservation is absent or nearly absent. How could it be?

Needless to add that it is a conserved vector current
that drives a dilepton production. In other words, the
absence of a structure seen in a dilepton production is
consistent with the chiral spin symmetry. 

\section{Review of the spatial correlators above $T_c$}

The spatial correlators of the isovector operators
$\mathcal{O}_\Gamma(\mathbf{0},0)$

\begin{equation}
C_\Gamma(z) = \sum\limits_{x, y, t}
\braket{\mathcal{O}_\Gamma(x,y,z,t)
\mathcal{O}_\Gamma(\mathbf{0},0)^\dagger}
\label{eq:momentumprojection}
\end{equation} 

\noindent
in $N_F=2$ QCD with the Domail Wall Dirac operator at physical
quark masses
have been calculated in Refs. \cite{R1,R2}. The isovector
fermion bilinears are named
according to Table \ref{tab:ops}. The change in names of bilinears
as compared to the temporal correlators is because a propagation
along the $t$-axis and along a spatial axis requires different
gamma-structures of operators.

\begin{table}
\center
\begin{tabular}{cccll}
\hline\hline
\rule{0pt}{3ex}
 Name        &
 Dirac structure &
 \quad Abbreviation  \quad &
 \multicolumn{2}{l}{
 } 
 
\\[1ex]
\hline
\rule{0pt}{3ex}

\textit{Pseudoscalar}        & $\gamma_5$                 & $PS$         &
 \multirow{2}{1cm}{$\left.\begin{aligned}\\ \end{aligned}\right] U(1)_A$}& \\
\textit{Scalar}              & $\mathds{1}$               & $S$        &  & 
\\[1ex]
\hline
\rule{0pt}{3ex}
\textit{Axial-vector}        & $\gamma_k\gamma_5$         & $\mathbf{A}$ & 
\multirow{2}{1cm}{$\left.\begin{aligned}\\ \end{aligned}\right] SU(2)_A$}&\\
\textit{Vector}              & $\gamma_k$                 & $\mathbf{V}$ & & \\
\textit{Tensor-vector}       & $\gamma_k\gamma_3$         & $\mathbf{T}$ & 
\multirow{2}{1cm}{$\left.\begin{aligned}\\ \end{aligned}\right] U(1)_A$} &\\
\textit{Axial-tensor-vector} & $\gamma_k\gamma_3\gamma_5$ & $\mathbf{X}$ & &\\[1ex]
\hline\hline
\end{tabular}
\caption{Fermion isovector bilinears and their $U(1)_A$ and $SU(2)_L \times SU(2)_R$ transformation properties
(last column). This classification assumes propagation in $z$-direction. The
open vector index $k$ here runs over the components $1,2,4$, i.e., $x,y$ and $t$.}
\label{tab:ops}
\end{table}

At finite temperature the rotational $SO(2)$ symmetry connects the 
correlators of the spatial components $V_x \leftrightarrow V_y$, $A_x \leftrightarrow A_y$ et cetera.   The CS-transformations (\ref{equ:su2cstrafos}) with $k=1,2$ together with the
$x \leftrightarrow y$ symmetry generate the following multiplets:
\begin{align}
(V_x,V_y); \; (A_x, A_y, T_t, X_t) \; ,  \label{equ:s2a} \\
(V_t); \;   (A_t, T_x, T_y, X_x, X_y) \; . \label{equ:s2b} 
\end{align}

\noindent
Extending the $SU(2)_{CS}$ group to $SU(4)$ one obtains larger 
multiplets of the isovector operators:
\begin{align}
(V_x, V_y, A_x, A_y, T_t, X_t) \; , \label{equ:ss2a} \\
(V_t, A_t, T_x, T_y, X_x, X_y) \; . \label{equ:ss2b} 
\end{align}
Complete  $SU(4)$ multiplets include also the isoscalar partners
of the operators $A_x, A_y, T_t, X_t$ in eq. (\ref{equ:ss2a})
as well as isoscalar partners of the $A_t, T_x, T_y, X_x, X_y$ operators
in eq. (\ref{equ:ss2b}).

\begin{figure}
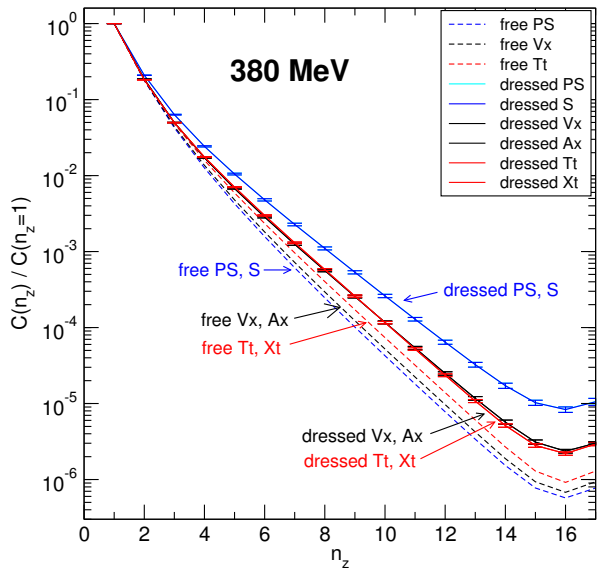

  \centering
  \includegraphics[scale=0.45]{{{4}}}
  \caption{Correlators of the $PS, S, V_x, A_x, T_t, X_t$ operators
     in full QCD at $T$= 380~MeV ($\sim 2.2 T_c$) for $8 \times 32^3$
     lattice (abbreviated as \textit{dressed})
      and with non-interacting quarks (\textit{free}) on the same lattice. From Ref. \cite{R1}.
  }
  \label{fig:e2}
\end{figure}

On Fig. \ref{fig:e2} 
we show correlators (full lines) calculated in  $N_F=2$ QCD at $T=380$ MeV ($\sim  2.2 T_c$)
normalized to 1 at $n_z=1$ where $n_z$ is the dimensionless distance in lattice
units \cite{R1}. The dashed lines represent calculation with  free
noninteracting quarks (i.e., they represent a QGP at an asymptotically
high temperature).

\begin{figure}
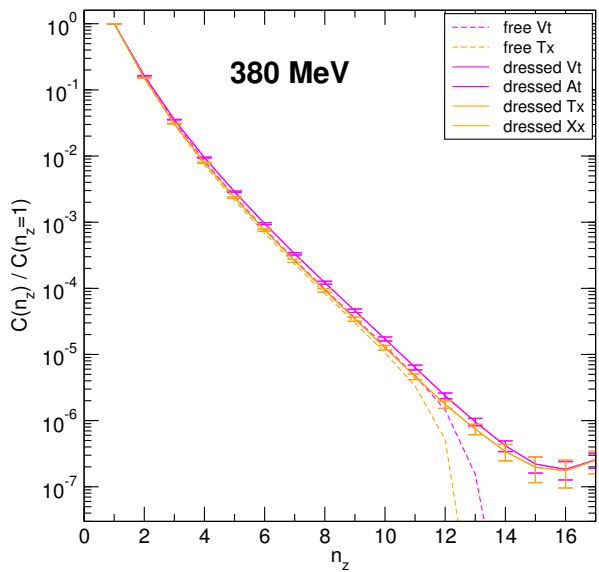

  \centering
  \includegraphics[scale=0.45]{{{5}}}
  \caption{Correlators of the $ V_t, A_t, T_x, X_x$ operators
     in full QCD at $T$= 380~MeV ($\sim 2.2 T_c$) for $8 \times 32^3$
     lattice (abbreviated as \textit{dressed})
      and with non-interacting quarks (\textit{free}) on the same lattice. From Ref. \cite{R1}. }
  \label{fig:e3}
\end{figure}

Degeneracy of the S- and PS-correlators indicates the $U(1)_A$
symmetry. The same is true for degenerate $T_t$ and $X_t$
correlators, that are connected by $U(1)_A$.
Equility of the $V_x$ and $A_x$ correlators is because of
the restored $SU(2)_L \times SU(2)_R$ symmetry.
An approximate degeneracy of the $(A_x, T_t, X_t)$
correlators evidences the $SU(2)_{CS}$ symmetry
while an
approximate degeneracy of all four correlators $(V_x, A_x, T_t, X_t)$
reflects the $SU(4)$ symmetry. 
This multiplet structure is very
different from the pattern of free correlators that demonstrates
only $U(1)_A$ and $SU(2)_L \times SU(2)_R$ symmetries.
 All these features have been discussed in detail in Refs. \cite{R1,R2}.

 On Fig. \ref{fig:e3} we show correlators normalized to 1 at $n_z=1$ 
 built with
the  $V_t, A_t, T_x, X_x$ operators.
A degeneracy according to the  multiplets (\ref{equ:s2b}) and(\ref{equ:ss2b}) is obvious.
Now comes an important point. The correlators of the $V_t$ and $A_t$
operators coincide with their free counterparts. This should not be true
 for operators $T_x$ and $X_x$. Their free correlators are different from
 the free correlators $V_t$ and $A_t$ exactly by the factor 2 
 as it follows from analytical continuum calculations \cite{R2}. The free correlators
 $T_x, X_x$ look degenerate with the free correlators $V_t, A_t$ only
 because of normalization of all correlators to 1 at $n_z=1$ 
 on Fig. \ref{fig:e3}\footnote{Note that while both dashed curves
 on Fig. \ref{fig:e3} must in reality exactly coincide after normalization to 1, they are slightly split only because of discretization errors on
 the lattice. Also their bending down around $n_z=12$ is a finite volume effect on the lattice.}
 
What distinguishes the  $V_t$ (and $A_t$) operator from the other ones
on Fig. \ref{fig:e2} and   Fig. \ref{fig:e3}? This operator is constrained by the current
conservation (\ref{current})\footnote{ The $V_x ~(A_x)$ and $V_y ~ (A_y)$ operators are transverse
to the propagation direction $z$ and hence are not constrained by (\ref{current}).}. 

Given results depicted on Fig \ref{fig:e2} and Fig. \ref{fig:e3}
as well as discussion above we conclude that similar to the temporal
correlators one observes multiplets of the chiral spin and $SU(4)$
groups and at the same time correlators constructed with operators that
are constrained by the $SU(2)_V$ current conservation show absence of
the interquark interaction.

The chiral spin and $SU(4)$ symmetries are not symmetries
of the Dirac action and are incompatible with free deconfined
quarks. Indeed, they are symmetries of the color charge (\ref{Qc})
and consequently indicate confinement. Deconfinement can happen
only when the color charge is screened and consequently the CS and
$SU(4)$ symmetries would disappear. But what does it mean that 
at the same time in  channels
that are constrained by a current conservation the temporal and spatial
correlators demonstrate a free-like behavior? We will answer this
question in the next section.

Now we will address fluctuations of conserved charges that
have been mentioned in Introduction. These fluctuations are
typically obtained via chemical potential derivatives of 
pressure \cite{Ejiri,Bazavov}. However they can also be obtained
in an alternative way.\footnote{The author is  thankful
 to T. Cohen for pointing this out and to F. Karsch for correspondence
 on this issue.}

The Lorentz-invariant
conserved quark (baryon) $U(1)_V$ charge $Q$
and flavor (isospin) $SU(N_F)_V$ charge  $Q^a_F$ are defined as 
spatial integrals of the corresponding charge densities:

\begin{equation}
Q \;  = \; \int \!\! dx dy dz  \; 
{\bar \psi}(x,y,z,t) \gamma^4 \psi(x,y,z,t),
\label{Q}
\end{equation}

\begin{equation}
Q_F^a \;  = \; \int \!\! dx dy dz  \; 
\bar{\psi}(x,y,z,t) \gamma^4 \frac{\tau^a}{2} \psi(x,y,z,t), ~~~ a = 1,2,...,N^2_F -1,
\label{Qi}
\end{equation}

\noindent
where $\tau^a$ are the isospin (flavor) generators.
The fluctuations of the conserved charges are completely determined by the spatial equal time
 correlators of the charge density operator $V_t$ for both flavor
 singlet and nonsinglet operators. This is because the quark
 (baryon) charge $Q$ (\ref{Q}) and the flavor charge $Q_F^a$ (\ref{Qi})
 are spatial integrals of the corresponding charge densities. Consequently
 the spatial equal time correlators of the quark (baryon) charge density and
 of the flavor charge density should contain complete information
 about fluctuations of conserved charges.

 The spatial correlator $V_t$ of conserved charge shown in Fig. \ref{fig:e3} (the flavor singlet correlator of the quark charge density, that is not shown on  Fig. \ref{fig:e3}, should be exactly the same) is identical with
 its free quark counterpart. In this correlator an integration over
 time is performed. From this equivalence it follows that the
 spatial equal time correlator in full QCD will also coincide with
 its free quark counterpart.
 Since the spatial equal time correlators contain complete information
 about fluctuations of  conserved charges it follows that  soon above $T_c$ the fluctuations of conserved charges should demonstrate absence of the
 interquark interactions, in agreement with Fig. \ref{fl}.
 
 Now we summarize both this and previous sections. While the temporal and spatial
 correlators above $T_c$ but below $3 T_c$ demonstrate clear patterns of
 the $SU(2)_{CS}$ and $SU(4)$ symmetries, which are symmetries of confinement
 and are incompatible with free quarks, at the same time these correlators
 show very soon above $T_c$ absence of the interquark interactions in 
 channels that are constrained by a current conservation. The latter
 channels are responsible for both dilepton production and for
 fluctuations of conserved charges. How to reconcile these seemingly contradicting facts? This will be discussed in the subsequent section.

\section{ Deconfinement in $SU(2)_c$ and confinement in $SU(3)_c/SU(2)_c$
in QCD}

We certainly do not know a microscopic mechanism that leads to
the present puzzling situation. However from the symmetry grounds we can obtain
some solid insights.

A confining interquark interaction in QCD persists until the color charge of quarks is not screened.
Once it is screened, then there is no color force between quarks and they behave as free particles. About existence of confinement in QCD at high
temperatures we can judge from the chiral spin  (and $SU(2N_F)$) symmetry of observables. They persist in the $T_c - 3 T_c$ interval \cite{R1,R2,R3}.
The chiral spin $SU(2)_{CS}$ symmetry is a symmetry in any $a=1,...,8$ component of the 
color charge (\ref{Qc}). Consequently observation of this symmetry
in the $T_c - 3 T_c$ interval implies that the color charge {\it is
not screened at least in some of the eight components}. Absence of
the strong interquark interactions in  channels constrained by a
current conservation requires that in these channels the color charge should
be screened. This could happen  if the color charge (\ref{Qc}) was screened
  in a $SU(2)_c $ subgroup of $SU(3)_c$.

 A strong color force is a consequence of a local gauge invariance.
 If the color charge is screened within a $SU(2)_c $, then the
 $SU(2)_c $ symmetry becomes global. A global $SU(2)_c $ symmetry implies via
 Noether theorem a conserved colored vector current,
 
\begin{equation}
\partial_\mu (\bar \psi \frac{t^a}{2} \gamma^\mu \psi) = 0,~~~ a=1,2,3.  
\label{colorcurrent} 
\end{equation}  
 
\noindent 
The conserved baryon charge and isospin current also satisfy a continuity
equation:

\begin{equation}
\partial_\mu (\bar \psi  \gamma^\mu \psi) = 0, 
\label{barcurrent} 
\end{equation}

\begin{equation}
\partial_\mu (\bar \psi \frac{\tau^a}{2} \gamma^\mu \psi) = 0,~~~ a=1,2,3.  
\label{isocurrent} 
\end{equation}  
 
\noindent
Then one needs a condition that would connect the color
current conservation within $SU(2)_c$ with the baryon (\ref{barcurrent})
and isospin current (\ref{isocurrent}) conservations.
Such a condition would be provided by the color - isospin locking,
i.e. by the locking of the color index of quarks with their
isospin index within a $SU(2)_c$ subgroup and such a locking
should be absent in  $SU(3)_c/SU(2)_c$.
This locking would automatically provide an equivalence
of  eq. (\ref{colorcurrent}) with eqs. (\ref{barcurrent})-(\ref{isocurrent}).
I.e. the absence of a strong confining interquark interaction 
happens   only in channels with conserved current. What a microscopical
reason for such locking could be is an open question.

This kind of locking supplemented with some dynamics 
was discussed in a different context for the color-flavor locking
at extremely high
baryon density \cite{A1,A2}. We stress that we use a locking only as
a mathematical device that allows to combine all facts discussed in this
paper. Its precise physical meaning should be clarified in subsequent
studies. 

It is not unreasonable to assume 
a locking soon after $T_c$. 
 The above scenario of the color-isospin locking guarantees that
 a deconfinement within a $SU(2)_c$ subgroup of $SU(3)_c$
 can be seen only in channels with conserved current.
 The eq. (\ref{colorcurrent}) is not true for all eight
generators of $SU(3)_c$. Consequently within the  $SU(3)_c$
confinement is still there.
 In this way we obtain simultaneously a chiral spin symmetry of correlators
 (i.e., confinement) and correlators of operators with a conserved current
 behave as if quarks were free. This means that the color-isospin
 locking filters out a $SU(2)_c$ subgroup and the $SU(3)_c/SU(2)_c$
 part of the color dynamics is not seen via vector and axial vector
 operators.

 Needless to add that this scenario implies that a dual superconductor
 picture of confinement in QCD \cite{Man,Hooft,SW}, that is an
 abelian $U(1)$ confinement, is quite far from  real life in QCD
 because it implies that above a critiacal temperature  quarks
 are free. In QCD there is still
 confinement in $SU(3)_c$.

 A general possibility that there  could be at high temperatures a partial 
deconfinement, i.e. that a $SU(M)$ subgroup of  the $SU(N_c)$,
$M < N_c$ is deconfined above $T_c$ was also discussed in Refs. 
\cite{H0,H1,H2,H3}.

\section{Conclusions}

We have reanalysed results on spatial and temporal
correlators above $T_c$ \cite{R1,R2,R3} and demonstrated
that while these correlators exhibit a chiral spin $SU(2)_{CS}$
and $SU(2N_F)$ symmetries at temperatures $T_c - 3 T_c$, they also
show absence of the interquark interactions in channels
constrained by a current conservation. The chiral spin and 
$SU(2N_F)$ are not symmetries of the Dirac action and hence
are incompatible with free deconfined quarks. They are symmetries
of the color charge in QCD and indicate that the theory is still
in the confining regime. 

Absence of the interquark interactions in channels with
a conserved current explains why fluctuations of conserved
charges soon after $T_c$ behave as if quarks were free. It
also explains why no bound states are seen via dileptons
above $T_c$.

In order to combine all these facts we have conjectured
that in the temperature region $T_c - 3 T_c$ a $SU(2)_{color}$
subgroup of $SU(3)_{color}$ is deconfined, i.e. a screening
happens in three components of the color charge, and five
generators of $SU(3)_{color}$ are still confined, i.e. not
screened. Assuming a $SU(2)_{color}$ - $SU(2)_{isospin}$ locking
in one of $SU(2)_{color}$ subgroups of $SU(3)_{color}$ we
explain why the chiral spin symmetry
persists and at the same time there is no $SU(2)_{color}$
color-mediated
strong force in channels with conserved current. 
These channels are still in a confining mode
since there is  confinement in $SU(3)_{color}/SU(2)_{color}$.

The author is grateful to Tom Cohen for many interesting
and stimulating discussions.



\begin{thebibliography}{99}

\bibitem{F1}
Y.~Aoki, S.~Borsanyi, S.~Durr, Z.~Fodor, S.~D.~Katz, S.~Krieg and K.~K.~Szabo,
   JHEP {\bf 0906}  088 (2009).

\bibitem{K}
A.~Bazavov {\it et al.} [HotQCD Collaboration],
  Phys.\ Lett.\ B {\bf 795}, 15 (2019).

\bibitem{eq1} 
S.~Borsanyi, Z.~Fodor, C.~Hoelbling, S.~D.~Katz, S.~Krieg and K.~K.~Szabo,
  Phys.\ Lett.\ B {\bf 730}, 99 (2014).

\bibitem{eq2} 
 A.~Bazavov, P.~Petreczky and J.~H.~Weber,
  Phys.\ Rev.\ D {\bf 97},  014510 (2018).

\bibitem{Koch} 
S.~Jeon and V.~Koch,
  Phys.\ Rev.\ Lett.\  {\bf 85}, 2076 (2000).

 \bibitem{Asakawa}
 M.~Asakawa, U.~W.~Heinz and B.~Muller,
  Phys.\ Rev.\ Lett.\  {\bf 85}, 2072 (2000).


\bibitem{Ejiri} 
F.~Karsch, S.~Ejiri and K.~Redlich,
  Nucl.\ Phys.\ A {\bf 774}, 619 (2006).

\bibitem{Bazavov}
A.~Bazavov {\it et al.},
  Phys.\ Rev.\ D {\bf 95}, no. 5, 054504 (2017).


\bibitem{R1} 
  C.~Rohrhofer, Y.~Aoki, G.~Cossu, H.~Fukaya, L.~Y.~Glozman, S.~Hashimoto, C.~B.~Lang and S.~Prelovsek,
  Phys.\ Rev.\ D {\bf 96}, 094501 (2017)
  Erratum: [Phys.\ Rev.\ D {\bf 99}, 039901 (2019)].

\bibitem{R2}
 C.~Rohrhofer, Y.~Aoki, G.~Cossu, H.~Fukaya, C. Gattringer, L.~Y.~Glozman, S.~Hashimoto, C.~B.~Lang and S.~Prelovsek,
   Phys.\ Rev.\ D {\bf 100}, 014502 (2019).
  

\bibitem{R3} 
C.~Rohrhofer, Y.~Aoki, L.~Y.~Glozman and S.~Hashimoto,
   Phys.\ Lett.\ B {\bf 802},  135245 (2020).

\bibitem{G1} 
  L.Y.~Glozman,
  Eur.\ Phys.\ J.\ A {\bf 51} 27 (2015).

\bibitem{G2}
  L.Y.~Glozman and M.~Pak,
  Phys.\ Rev.\ D {\bf 92}, 016001 (2015). 

\bibitem{G3}
  L.Y.~Glozman,
  Eur.\ Phys.\ J.\ A {\bf 54}, 117 (2018).

\bibitem{G4}
 L.~Y.~Glozman,
  arXiv:2004.07525 [hep-ph].



 \bibitem{CME1}
D.~E.~Kharzeev, L.~D.~McLerran and H.~J.~Warringa,
  Nucl.\ Phys.\ A {\bf 803}, 227 (2008).


\bibitem{CME2}
K.~Fukushima, D.~E.~Kharzeev and H.~J.~Warringa,
  Phys.\ Rev.\ D {\bf 78},  074033 (2008).
 
\bibitem{CME} 
 J.~Zhao [STAR Collaboration],
  arXiv:2002.09410 [nucl-ex].

\bibitem{D1}
  M.~Denissenya, L.Y.~Glozman and C.B.~Lang,
  Phys.\ Rev.\ D {\bf 89} 077502 (2014).

\bibitem{D2}
M.~Denissenya, L.~Y.~Glozman and C.~B.~Lang,
  Phys.\ Rev.\ D {\bf 91}, 034505 (2015).

\bibitem{Tomiya:2016jwr}
  A.~Tomiya, G.~Cossu, S.~Aoki, H.~Fukaya, S.~Hashimoto, T.~Kaneko and J.~Noaki,
  Phys.\ Rev.\ D {\bf 96}, no. 3, 034509 (2017)
  Addendum: [Phys.\ Rev.\ D {\bf 96} 079902 (2017)].

\bibitem{A1}
 M.~G.~Alford, K.~Rajagopal and F.~Wilczek,
  Phys.\ Lett.\ B {\bf 422}, 247 (1998).

\bibitem{A2}
M.~G.~Alford, A.~Schmitt, K.~Rajagopal and T.~Schäfer,
  Rev.\ Mod.\ Phys.\  {\bf 80}, 1455 (2008).

\bibitem{Man} 
 S.~Mandelstam,
  Phys.\ Rept.\  {\bf 23}, 245 (1976).

\bibitem{Hooft} 
 G.~'t Hooft,
  Nucl.\ Phys.\ B {\bf 153}, 141 (1979).

\bibitem{SW} 
 N.~Seiberg and E.~Witten,
  Nucl.\ Phys.\ B {\bf 426}, 19 (1994)
  Erratum: [Nucl.\ Phys.\ B {\bf 430}, 485 (1994)].


\bibitem{H0}
 M.~Hanada, G.~Ishiki and H.~Watanabe,
  JHEP {\bf 1903}, 145 (2019)
  Erratum: [JHEP {\bf 1910}, 029 (2019)].

\bibitem{H1}
 M.~Hanada, A.~Jevicki, C.~Peng and N.~Wintergerst,
  JHEP {\bf 1912}, 167 (2019).
  
\bibitem{H2} 
  M.~Hanada, G.~Ishiki and H.~Watanabe,
  arXiv:1911.11465 [hep-lat].


\bibitem{H3}
 H.~Watanabe, G.~Bergner, N.~Bodendorfer, S.~Shiba Funai, M.~Hanada, E.~Rinaldi, A.~Schäfer and P.~Vranas,
  arXiv:2005.04103 [hep-th].

\end{thebibliography}
\end{document}